\documentclass[10pt,prb,aps,twocolumn,showpacs,amssymb,superscriptaddress]{revtex4}
\usepackage{graphicx, bm, amsmath, amsfonts}
\usepackage{color}

\newcommand{\SRCUBO}{ SrCu$_2$(BO$_3$)$_2$ }

\begin{document}

\title{Study of the Shastry Sutherland Model Using Muiti-scale Entanglement Renormalization Ansatz}

\author{Jie Lou}
\affiliation{Department of physics, Fudan University, Shanghai, China 200433}

\author{Takafumi Suzuki}
\affiliation{Research center for nano-micro structure science and engineering, Graduate school for engineering, Univiersity of Hyogo, Himeji, Japan 671-2280}

\author{Kenji Harada}
\affiliation{Graduate school of informatics, Kyoto University, Kyoto, Japan 615-8063}

\author{Naoki Kawashima}
\affiliation{ISSP, University of Tokyo,  
Kashiwa 5-1-5, Kashiwa, Japan 277-8581}

\begin{abstract}
We performed variational calculation based on the multi-scale entanglemnt renormalization ansatz,
for the antiferromagnetic Heisenberg model on a Shastry Sutherland lattice (SSL). 
Our results show that at coupling ratio $J'/J=0.687(3)$,
the system undergoes a quantum phase transition from the orthogonal dimer order to 
the plaquette valence bond solid phase, which then transits into the antiferromagnetic order above $J'/J=0.75$. 
In the presence of an external magnetic field, our calculations show clear evidences 
of various magnetic plateaux in systems with different coupling ratios 
range from $0.5$ to $0.69$. 
Our calculations are not limited to the small coupling ratio region,
and we are able to show strong evidence of the presence of several supersolid phases, 
including ones above $1/2$ and $1/3$ plateaux. 
Such supersolid phases,  
which feature the coexistence of compressible superfluidity and crystalline long range order
in triplet excitations,
emerge at relatively large coupling ratio ($J'/J>0.5$). 
A schematic phase diagram of the SSL model in the presence of magnetic field is provided.
\end{abstract}

\date{\today}

\pacs{75.10.Jm, 75.10.Nr, 75.40.Mg, 75.40.Cx}

\maketitle
%%%%%%%%%%%%%%%%%%%%%%%%%%%%%%%%%%%%%%%%%%%%%
%\section{Introduction and Highlight}
\section{Introduction}
\label{sec:INTRO}
%%%%%%%%%%%%%%%%%%%%%%%%%%%%%%%%%%%%%%%%%%%%%
%%edited by TS !!Note In related papers of supersolid physics, authors often use abbreviation SS for supersolid. So to avoid confusion, I changed Shastry-Sutherland lattice from SS to SSL. 
Geometrically frustrated interactions and quantum effect often lead to exotic orders such as, quantum spin ice on a pyrochlore lattice,\cite{Onoda} spin liquid state on Kagom\'e lattice\cite{Kagome} and fractionalized magnetization plateaux,\cite{Sebastian2008} because strong fluctuations suppress stabilization of classical or trivial magnetic orderings.
A Shastry-Sutherland lattice (SSL)\cite{Shastry} is one of such frustrated systems and the Hamiltonian of the SSL model can be written as
\begin{eqnarray}
H=J\sum_{\langle i,j \rangle} {\bf S}_i \cdot {\bf S}_j + J'\sum_{\langle \langle i,j \rangle \rangle} 
{\bf S}_i \cdot {\bf S}_j - h\sum_{i} S^{z}_i,
\label{Hamiltonian}
\end{eqnarray}
where $\langle i,j \rangle$ and $\langle \langle i,j \rangle \rangle$ stand for intra-dimer and 
inter-dimer site-pairs on the SSL as shown in Fig.~\ref{mera-structure}, panel (d).
%\comment{Kenji: please see comments in Fig.~\ref{mera-structure}.}
\SRCUBO has provided a good stage for studying the magnetic properties of the SSL model.\cite{Kageyama}
The crystal structure of this compound is characterized by a layered structure of ${\rm CuBO_{3}}$ and ${\rm Cu^{2+}}$ ions carry $S=1/2$ spins.\cite{SrCu2O3_structure}
From magnetic susceptibility measurements,\cite{SrCu2BO3_susceptibility} it has been reported that the ground state is spin gapped and the coupling ratio is estimated as $J'/J\sim 0.635$.\cite{Ueda-xi}
This compound shows anomalous properties in magnetic fields owing to the magnetic frustration; the magnetization curve exhibits a multi-steps structure.
In addition to the plateaux with magnetization at $1/8$, $1/4$ and $1/3$ of the saturation value,\cite{Takigawa,experiments} recent torque experiments suggested existence of $1/6$ and other plateaux with more exotic fraction numbers.\cite{Sebastian, Jaime}

%%% It may be better to move to summary.
Quite recently, rare-earth compounds\cite{TbB4, TmB4_1,TmB4_2,TmB4_3,Yb2Pt2Pb} with the SSL structure were also studied in experiments, 
showing exciting magnetic properties may or may not differ from the original material \SRCUBO.
While the inter-dimer and intra-dimer interaction strength differs from material to material and long range interactions can be crucial to understand magnetic properties in the rare-earth compound cases, the SSL model with frustrated Heisenberg interactions is still the key point to understand the physics of magnetic properties of these materials.

Theoretically, the ground-state phase diagram of the Hamiltonian (\ref{Hamiltonian}) has also been in debate for this decade. 
The remarkable feature is that, in the limit $J'/J \rightarrow 0$, the direct product state of singlets on $J'$ bonds becomes the eigen state and the energy of this orthogonal dimer (OD) state can be given exactly by $E=-3/8J$. 
Oppositely, the system becomes equivalent to the antiferromagnetic (AF) Heisenberg model on a square lattice in the limit $J'/J \rightarrow \infty$. 
Thus the conventional AF state is realized with a gapless dispersion. 
This fact leads to naive expectation that a quantum phase transition between these two phases take place when $J'/J$ is changed. 
Although the quantum phase transition including the possibility of intermediate phase has been intensively studied by analytical 
and numerical approaches,\cite{Ising_expansion,dimer_expansion,plaquette,Ueda-ED,exactdiag,schwingerboson,seriesexpansion,PEPS} conclusive result has not been established yet. 

In fact, obtained results strongly depend on approaches.
More specifically, the high-temperature-series expansion, early exact diagonalization and PEPS\cite{Ueda-ED,seriesexpansion,dimer_expansion,PEPS} suggested the direct transition from the OD phase to the AF phase. 
In contrast, Schwinger boson meanfield theory suggested the existence of helical ordered state as an intermediate phase.\cite{schwingerboson} Koga and Kawakami predicted that the other spin gapped phase exists for $0.677<J'/J<0.86$.\cite{plaquette} and it can be expressed by the alignment of resonating singlet dimers on the plaquette without $J$ bond, namely plaquette singlets. This prediction was supported by the other numerical exact diagonalization study on up-to 32-site systems.\cite{exactdiag}
More fascinating possibility was predicted from the results of dimer-field theory in ref.~\onlinecite{DFT}; weakly incommensurate spin-density wave state is stabilized in the intermediate phase. 
The modified cluster expansion with self-consistent perturbation by Hajj and Malrieu suggested that the lowest energy state is either plaquette singlet or columnar dimer states.\cite{SCP}
The reason of the above long historical discussions on the intermediate region is caused by the fact that quantum fluctuation leads to numerous candidate states, all of which have similar energies. 
Therefore one may hope unbiased calculations such as quantum Monte Carlo (QMC). Since QMC calculation, which is very helpful to settle problems if possible, suffers from the negative sign problem in the present case, this approach will be excluded.

Similar situation is also seen in magnetic properties when external field $h$ exists: multi-plateau physics.
%% until here
Exact diagonalizations\cite{Ueda-ED} and other pioneer studies\cite{Momoi} have established the existence of 
$1/2$, $1/3$ and $1/4$ plateaux, while details of the small magnetization region were not very clear.
In a recent paper by Sebastian {\it et al.},\cite{Sebastian} the possibility of fractional magnetic plateaux at $1/q$ ($q$ = integer) was discussed in analogy to the quantum Hall effect. 
Several authors predicted many possible plateaux with further exotic fraction numbers like such as 
$1/9$ and $2/15$.\cite{Capponi,Mila-plateaux}
They focused on the strong dimer limit $J' \ll J$ case, and inter-dimer interactions $J'$ were treated as perturbations.\cite{Momoi, Mila-plateaux}
In such case, triplet excitations induced by external magnetic fields can be regarded as hardcore bosons and the interactions on plaquette without $J$ bonds can be considered as a combination of repulsion interaction and hopping process.\cite{Momoi} 
Although many believes that these plateaux shall persist in the large coupling ratio ($J'/J>0.5$) region, up to best of our knowledge, 
no unbiased numerical method has thoroughly confirmed the result in such situations.

What is more interesting is the proposal of possible supersolid phases which may appear in the presence of large external field.\cite{Momoi}
The supersolid phases have been confirmed in various 1D and 2D frustrated models,\cite{supersolid} and it has been well established that correlated hopping 
(bosons hop when another boson is present as its neighbor)
plays an important role,\cite{Mila-supersolid}
since single boson hopping is strongly suppressed by frustration in the SSL model.
Due to the strong competition between repulsion and hopping terms,
supersolid phases are unlikely to emerge as the ground state at small coupling ratio region ($J'/J<0.5$).
As a result, supersolidty has not been well investigated in early calculations which treat $J'$ as simple perturbation.\cite{Capponi,Mila-plateaux}

In this paper, we study the magnetic properties of the SSL model,
using variational method based on the entanglement renormailzation (ER) ansatz. 
This method is more popularly known as ``multi-scale entanglement renormalization ansatz'', 
or simply MERA, which is proposed by Vidal a few years ago.\cite{Vidal}
It is a promising numerical method based on merging the idea of tensor product state ansatz 
and renormalization group theory.
The key feature of MERA is the introduction of so called ``disentanglers'' which targets on reducing local correlations, so that long range entanglement can be captures with limited computational cost at a coarse grained level. 
In our ER based calculation however, we simply use one level of coarse graining, 
instead of multiple levels in MERA calculations.
We employ multiple neighboring unitcells in neighborhood of the unitcell we calculate,
to study thermodynamic properties with meanfield approximation.
This method is also proposed by Evenbly and Vidal under the name ``finite range MERA'' (FRMERA),\cite{Vidal-FR}
due to the fact that only entanglement within finite range up to the size of each 
unticell is well captured.
We will discuss details of FRMERA and justify the reason we use this method in our ER calculation.

We want to highlight that unlike previous studies mentioned above, {\it our calculation does not enforce any bias to the original Hamiltonian}.
The ground state is obtained from pure variational calculations, as the state with lowest energy for the input Hamiltonian ({\it i.e.}, we provide upper boundary for the ground state energy.)
Our calculation focuses on the relatively large coupling ratio region, which has not been well investigated, since we are not limited to the small coupling ratio limit $J' \ll J$. 

In the absence of magnetic field, we find strong evidence of the existence of the plaquette valence bond solid (VBS) phase in the intermediate coupling ratio region above the first phase transition point $J'/J=0.687(3)$.
When the external field is turned on, we confirm the all the major (well accepted) plateaux such as $1/4$, $1/3$ and $1/2$, as well as minor ones including $1/6$, $1/8$ and $1/9$.
We also observe a strong evidence of several supersolid orders above $1/3$ and $1/2$ plateaux, as
proposed by Momoi {\it et al.},\cite{Momoi} which has not been discussed in previous numerical studies to best of our knowledge. 
Our results show that the $1/3$ plateau only appears above $J'/J=0.5$, and the $1/2$ plateau and its corresponding supersolid phase persist until the largest coupling ratio $J'/J=0.687$ before the phase transition to the plaquette order.
Our schematic phase diagram is shown in Fig.~\ref{phase-diagram}, Sec.~\ref{sec:MP}. 

The remaining parts of this paper is organized as following:
In Sec.~\ref{sec:MERA}, we discuss the ER method and tensor network structure we designed for the SSL model.
In Sec.~\ref{sec:GS}, results of the ground state in {\it absence} of magnetic field are presented.
We show magnetic behaviors of the SSL model from our calculation, including plateaux and supersolid in Sec.~\ref{sec:MP}.
Discussion and summary are presented at last.
%%%%%%%%%%%%%%%%%%%%%%%%%%%%%%%%%%%%%%%%%%%%%
\section{MERA method and tensor network structure}
\label{sec:MERA}
%%%%%%%%%%%%%%%%%%%%%%%%%%%%%%%%%%%%%%%%%%%%%
%--------------------------------------------------------------------------
\begin{figure}
\centerline{\includegraphics[width=7.5cm,clip]{structure_mera.epsi}}
\caption{Illustration of tensor netwrok structure used in ER calculation.\\
Panel (a): The shastry sutherand model plotted on a square lattice.
Intra-dimer (J) and inter-dimer (J') couplings are shown as solid and dashed line,
respectively.
In the entanglement renormalization procedure,
8 original spins included in each domain (shown as shadow region) are coarse grained into one
(shown as blue dot) by a blue isometry tensor.
Two sets of disentanglers (red and green squares) are applied on boundaries between different domains.
\\
Panel (b): two sets of disentangler tensors and one set of isometry tensors are applied 
in a consecutive manner,
following the coarse graining direction  
from bottom (starting hamiltonian) to top (coarse grained system).
Lower and upper legs stand for in and out indices, respectively.\\
Panel (c): a final density matrix is applied on the coarse grained system (indicated by blue circles).
For an unitcell (denoted by shadows in the background) with $4\times 6$ spins, 
3 coarse grained sites are involved in the density matrix,
which is indicated by a light triangle.
For unitcell shaped $6 \times 8$, 6 coasre grained sites are inculded,  
by doubling the length along the y direction. 
In FRMERA calculation, neighboring unitcells (shadow regions) are used as 
meanfield environment to simulate system in the thermodynamic limit $L \rightarrow \infty$.
\\
Panel (d): 
%Overviewing the tensor network from above (CG level), all interactions are covered by tensors.
%A unitcell with $32$ sites which preserve $\pi/2$ rotational symmetry of the square lattice is show%n as shaded region.\\
Unitcells consist of $4\times 8$ spins, or equivalently 4 coarse grained sites.
This unitcell is specially chosed to perserve the X-Y symmetry of the square lattice.
%\comment{Kenji: can you show the J and J' bonds of the SSL model in panel (a)? Is it better that panel (d) shows the unitcell of 32 sites. If we use the unitcell of 48 sites in Sec.~\ref{sec:MP}, panel (e) or an additional explanation in panel (c) is useful.}\\
%Jie: modeifed according to Harada san's suggestion.
}
\label{mera-structure}
\end{figure}
%--------------------------------------------------------------------------

The structure of tensor network plays an important role in ER calculations. 
Although there are no strict rules for constructing ER tensor network for a specific model, one guideline shall be taken into consideration. 
The advantage of ER based method comes from the {\it disentangler} tensors, which reduce local correlations so that the density matrix of the coarse grained (CG) system is capable of capturing long range entanglement of the original spin model with relatively small virtual dimensions $\chi$.
Naturally, {\it all interactions across boundary of the coarse graining domain should be treated by disentangler tensors}. 
%{\bf otherwise (idea for a better sentence?)} 
%Failed to do so, short range entanglements induced from such 
%leftover interactions will require
%considerably larger dimension $\chi$ on the CG level,
%in order to capture these short range entanglements,
%thus make the calculation much more demanding.
%Naively, one can introduce disentangler tensors large enough to cover all such interactions.
%Alternatively include many layers of tensors. 
%Unfortunately, both approaches increase computational cost rapidly, so compromise has to be made.
%It is especially important to cover interactions on the boundarie between two adjacent coarse graining domains.
Relatively speaking, in most frustrated two dimensional quantum spin systems, it is common to have more interactions involved with each site, compared to non-frustration spin models.
As a result it is important to design an ER tensor network suitable for the SSL model that we are interested in.

In Fig.~\ref{mera-structure}, we show an illustration of the ER tensor network structure we chose for this calculation.
Notice that the network is constructed on the square lattice's geometry, instead of the SSL's.
We believe that it does not affect the result much in the small coupling ratio limit, as the ground state can be considers as decoupled dimers separated on the lattice.
Moreover, such a choice may aid us in the large coupling ratio region where we are most interested in.
This geometry allows us to easily utilize our tensor structure to study other frustrated lattice
which can be mapped onto square lattice ({\it i.e.}, the checker-board model).

Overall, 8 sites on the original square lattice are coarse grained into one site on the renormalized lattice, which is another square lattice tilted by $\pi/4$.
Coarse-graining sites obtained from isometry tensors and the corresponding domain containing the 8 original spins are shown as blue circles and shadowed regions, respectively, in Fig.~\ref{mera-structure}, panel (a).
Two layers of 4-sites disentanglers (colored by red and green) are applied consecutively on the boundary of coarse-graining domains, before the isometry tensors are applied. 
%Based on panel (d),\comment{If we change the figure, the reference should be changed.} 
Overlook the tensor network on top of the SSL, it is clear that all intra and inter dimer interactions are covered by at least one tensor of the network.
As a result, all interactions across the boundary of CG domains are treated by disentanglers.

The coarse grained lattice is then described by an top density matrix.
Obviously, the dimension of the coarse-grained spin $\chi$ and the number of CG spins covered by the density matrix determine computational cost of our calculation.
In practice , we are able to treat up to 6 coarse grained sites ($4\times 8$ spins) per unitcell.

As mentioned previously, unlike the MERA method, 
we only apply one level of renormalization to the SSL in our calculation.
Instead of limiting ourselves to finite-size systems with periodic boundary conditions, we choose FRMERA to construct multiple identical neighboring unitcells. 
%along with the original one, all sharing same sets of tensors and network structure, to simulate the infinite lattice.
In Panel (c) of Fig.~\ref{mera-structure},
we illustrate an example when the unitcell contains $4\times6 = 24$ spins (shadowed region in the background).
In such a case, the density matrix of each unitcell (represented by light triangles in Fig.~\ref{mera-structure}) is not correlated to its neighbors at all. %({\it i.e., the virtual links connect those } 
As a result, the FRMERA method is obviously related to the meanfield approximation in the way how we simulate system in thermodynamic limit.

The reason we choose FRMERA instead of the ordinary MERA is twofold.
First, although applying multiple levels of renormailzation can significantly increase the system size we can study in MERA, associated cost is very high.
This is due to the fact that sufficiently large virtual dimensions $\chi$ is required for each site in the coarse grained system in order to capture sufficient long range entanglement in higher level of renormalization.
Moreover, we are not specially focusing on quantum critical region whereas scale-invariant physics play an important role (where Multi-level renormailzation is important and meaningful).
Rather, we pay special attentions to the ground state of SSL model with and without presence of external magnetic field.
In such a case, as long as the unitcell is large enough to accommodate candidate ground states and its associated entanglement, we can simply use ER with meanfield approximation which is significantly cheaper in computational cost compared with MERA.

As a matter of fact, FRMERA turns out to be beneficial in studying the magnetic properties of SSL model.
The SSL model shows intriguing magnetic properties in response to external magnetic field $h$.
Besides $1/3, 1/4$ and $1/8$ magnetic plateaux observed in early experiments, there are theoretical studies as well as possible signature from experiment supporting existence from plateaux with small fraction numbers, such as $1/6$, $1/9$ and $2/15$.\cite{Sebastian,Capponi}
One difficulty to study these magnetic plateaux computationally lies in the fact that they only appear as ground states if the unitcell itself is compatible with the fraction number.
For example, plateaux with fraction number $1/3$, $1/6$ and so on can not be well described in ER calculations performed on $4\times 8$ unitcell, nor can it be even with multiple renormalization applied (due to periodicity).
Obviously, it is increasingly difficult to capture plateaux with even smaller fraction numbers, which require larger unitcells to settle in.
The FRMERA method offers a semi-solution to this problem. 
Namely, one can further sort unitcells into different categories, each contains a different set of tensors.
Following this, we can effectively create a larger super-unitcell, which combines one unitcell per category, at a computational cost linear to the size (number of unitcells included). 
we show such super-unitcells as background in Fig.~\ref{pattern}.
Notice that unitcells shown in different colors belong to different categories.
%each has its own set of tensors.
As one example, in the last panel, the super-unitcell is considered as a combination of 3 unitcells, one per each color, so that the ER calculation is capable to capture the $1/9$ plateau.

It is worth mentioning that unitcells included in the super-unitcell do not have correlations with each other {\it at top level}, since it is just a product state like a meanfield approximation.
On the other hand, correlations can still pass through in lower level tensor network, so that we are still able to study plateaux with small fraction numbers which would be impossible to find in ordinary unitcell, or naively increase number of renormalization steps.

%Similar to other numerical methods (DMRG, PEPS, TERG) based on the tensor product state ansatz,
%the accuracy %(and the validity of the ground state found) 
%of results in MERA or FRMERA is largely dependent on
%the dimension of virtual indices linking tensors. 

%\comment{Jie: is this discussion needed? I think maybe it is not a good idea to say so: 
%In DMRG and PEPS, a cutoff dimension D$_{cut}$ is used as a compromise
%to computational cost, which induce approximations in the calculation. 
%In contrast, MERA is a pure variational method, and we provide an upper bound for the ground state
%energy.}\comment{Kenji: I only remain the values of virtual dimensions}
The dimension of coarse grained virtual spin, $\chi$, affects our
calculation results.
For unitcells consists of 3, 4 and 6 coarse grained spins per density matrix,
(which correspond to 24, 32, 48 original spins,)
we are able to perform calculations up to $\chi=9, 9$ and $4$, respectively.
It is necessary and useful to check the convergence (or evolution trend) 
of physical quantities as a function of the dimension $\chi$.   
We will discuss behaviors of evolution trend when needed in following discussions.

%Although the MERA structure we use has inherent geometry advantage to describe the SS model,
%there is one shortcome, namely, a two body $J$ interaction may end up in the coarse grained 
%level as a six body interaction (in the worst case). 
%As a result, it is difficult to apply multiple level of renormalization to study large
%lattice, and we are limited to a system as large as $6 \times 8$ sites.
%To overcome such problem, we use the finite range MERA (FRMERA), 
%proposed by Vidal,\cite{Vidal-FR}
%which in principle allow us to study infinite lattice, with one layer of renormalization. 
%The key point is to  
%instead of using conventional periodic boundary condition.
%At the top level, The wave function of the full system is considered to be a direct product of 
%all unitcells, ({\it i.e.}, $\chi_{top}=1$), instead of a fully entangled state which can only 
%be described with $\chi_{top}\geq 2$.
%In another word, we only focus on the calculation of one unitcell, and treat its neighbors
%as meanfield approximation.
%One example of FRMERA structure consists of $4\times 6 $ sites unitcells 
%is shown in Fig.~\ref{mera-structure}, panel (c),
%in which shadowed regions in the background represent all the unitcells mentioned above. 

%%%%%%%%%%%%%%%%%%% modified by T. Suzuki 2012.10.24 From here

%%%%%%%%%%%%%%%%%%%%%%%%%%%%%%%%%%%%%%%%%%%%%
\section{Ground state of SSL model in absence of external field}
\label{sec:GS}
%%%%%%%%%%%%%%%%%%%%%%%%%%%%%%%%%%%%%%%%%%%%%

%--------------------------------------------------------------------------
\begin{figure}
\centerline{\includegraphics[width=7.5cm,clip]{ground_state.eps}}
\centerline{\includegraphics[width=4cm,clip]{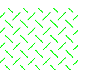}
\includegraphics[width=4cm,clip]{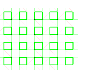}}
\caption{Upper panel: AF and dimer order parameters, $M$ and $D$, as a function of coupling ratio $J'/J$.
Results obtained from unitcell sized $4\times 6$ and $4\times 8$ are shown in black and red (light) colors,
respectively.
A strong first order transition occur at $J'/J=0.687(3)$, where the system enter the plaquette
VBS order which features finite dimer order parameter $D$. 
The transition point 
between the plaquette order and the standard AF order is not clear,
as the crossing point between $M$ and $D$ shifts significantly as the unitcell size is increased.
Energies of ground state (normalized by $J'$)
%\comment{Kenji: $J$? Jie: J' is correct.} 
are shown in the inset, with dashed lines
plotted as ``guide for eyes'' for phase boundaries. 
Energies obtained from larger unitcell $4\times 8$ is 
slightly better compared with those obtained for $4 \times 6$ unitcell.
%\textcolor{red}{In legend, red squares are for N=32? Y label in the inset should be $E/J'$.}
Lower panel: local bond energy profile for the OD order and the VBS order.
Thickness of the bond corresponds to the correlation between two spins $\langle S_i \cdot S_j \rangle$.
}
\label{ground-state}
\end{figure}
%--------------------------------------------------------------------------

Without external magnetic field, the true ground state of the Heisenberg model on the SSL remains to be controversial, but there are two trivial reference limits for the ground state.
In the limit $J' \ll J$, the system can simply be considered as a product of decoupled singlet dimers, or commonly referred to as the OD model. 
In another limit, the system reverts back to the non-frustrated Heisenberg model on the square lattice, with the AF order being the ground state.
The debate is mainly focused on the intermediate region where quantum fluctuations are large and sign-problem due to frustration of SSL model prohibits effective numerical methods for quantum magnets to be performed. 
%To study the intermediate region, we calculate the ground state of this model by a MERA tensor network. 
%As pointed out in Sec. II, our tensor network is compatible with candidate ground states of the SSL model. 
In Fig.~\ref{ground-state}, we show energy (inset) and order parameters as a function of coupling ratio $J'/J$.

To discuss the possibility of intermediate phase, namely plaquette phase, we introduce two estimators as the order parameters.
The ordinary AF order parameter $M$ is defined as
\eqnarray
M&=& \sqrt{M_x^2+M_y^2+M_z^2}, \\
M_x&=& \frac{1}{N} \sum_i \langle S^x \rangle _i (-1)^{x_i+y_i}, \\ 
\endeqnarray
and $M_y$, $M_z$ are defined likewise.
%\mbox{\comment{Kenji: is it correct?}}\\
The order parameter $D$ for the valence bond solid order is defined as
\eqnarray
D_{x}&=& \frac{1}{N} \sum_i \langle {\bf S}_i \cdot {\bf S}_{i+\hat{x}} \rangle (-1)^{x_i},\\
D_{y}&=& \frac{1}{N} \sum_i \langle {\bf S}_i \cdot {\bf S}_{i+\hat{y}} \rangle (-1)^{y_i},\\
D&=& (D^2_x+D^2_y)^{\frac{1}{2}},
\endeqnarray
where $x_i$, $y_i$ are coordinates of site i,
and $\langle i,j \rangle$ stands for nearest neighbor pairs on the square lattice.
%\comment{Jie: for M, The $z$ of $S_z$ means the direction of ordered magnetization, 
%it can be any direction due to su(2) symmetry. How to explain this or leave it simply like that?}
%For the dimer order parameter, $D_y$ is defined in the same form as $D_x$.
$D_{x}$ and $D_{y}$ are not direct representations of plaquette state, but they are ideal estimators because we can distinguish columnar and plaquette VBS order by evaluating the difference between $x$ and $y$ components of the dimer order, $D_{\Delta}= |  |D_x| - |D_y| |$, which is also shown in Fig.~\ref{ground-state}.

It is clear that the system is in the OD phase in the small coupling ratio region, because the energy can be given exactly by $E=-3J/8$. 
In fact, AF and dimer order parameters are strictly zero, and the energy of the system is linear to $J'$.

As $J'/J$ increases, %the system show a clear phase transition from the OD phase.
the ground state energy show a cusp at $J'/J=0.687$, and the dimer order parameter $D$ jumps to a finite value keeping the value $D_{\Delta} \sim 0$. (see red squares and diamonds in Fig.~\ref{ground-state}.) 
The system clearly undergoes a first order phase transition from the OD phase into the plaquette VBS phase. 
The estimated transition point is in good agreement with that in previous works.
As further evidence of the plaquette VBS order, we evaluate the local bonding energy and show the pattern in the lower panels of Fig.~\ref{ground-state}.
The bond energy profile also strongly indicates that the ground state in the intermediate region does not break $\pi/2$ rotational symmetry.

Results of FRMERA calculation are dependent to the size of unitcell we use. 
In Fig.~\ref{ground-state}, we show two sets of data which correspond to calculations based on two types of unitcells sized $4\times 6$ and $4\times 8$.
The dimer order parameter $D$ increases when the unitcell is enlarged.
The $4\times 6$ unitcell clearly breaks the rotational symmetry of the original model, whereas the fan-shaped $4\times 8$ unitcell does not (both shown in Fig.~\ref{mera-structure}).
As a result, in the intermediate region just above $J'/J>0.687$,
$D_{\Delta}$ remains small but finite in $4\times 6$ unitcells' calculations,
but turns out to be $0$ in the $4\times 8$ sized unitcells' case.
%\comment{Kenji: in Fig.~\ref{ground-state}, is it not correctly 0? Jie: there is a very small number, maybe some finite size effect?}
We conclude that plaquette VBS state, which preserve $\pi/2$ rotational symmetry of the square lattice, emerges as the ground state when the system undergoes a phase transition out of the OD order.

In contrast to the phase transition point between the OD phase and plaquette VBS phase, the transition between the AF order and plaquette VBS phase seems not to be clear due to finite size effect of our unitcell.
The value of $D$ drastically decreases around $J'/J \sim 0.75$ and $M$ starts to develops inversely.
However, the crossing point of two order parameters is shifting to higher $J'/J$ value, when the size of the unitcell is increased.
In fact, order parameters are slightly affected by dimension $\chi$ ({\it i.e.}, not fully converged) for $0.85 \agt J'/J \agt 0.75$.
Moreover, both order parameters remain finite in a relatively large region near the transition point.
As a result, solely based on available result, we can not clearly determine the nature of the phase transition between the plaquette VBS order and the AF order.
Further calculation is required for precise discussion of the phase transition point.

It is necessary to discuss the validity of the ground state we obtained from ER calculation.
First, it is natural to expect that MERA and ER methods are biased {\it in the sense that the tensor network (including the way we choose coarse grained lattice) has a spacial structure}.
Actually, the ER (and MERA) procedure probably breaks spacial (translational and rotational) symmetries inhering in the original Hamiltonian.
As a result, optimized wave functions of the ground state may contain unwanted effective perturbation introduced from the tensor structure.
Naturally, if the dimension $\chi$ of coarse grained system is sufficiently large, the effect of such bias can be neglected, and the solution of the ER calculation becomes exact. 
In practice, $\chi$ used in our calculation is limited, to avoid explosion of simulation cost. 
Such limited $\chi$ will cause severe problem to determine the nature of the ground state in the critical region, or to locate the phase transition point, where the gap between energies of different candidate orders are closing. 

What we can do is to reduce such bias by choosing an ER tensor structure that is compatible to most candidates of ground states.
For the present case, we specifically chosen a tensor network structure with even periodicity ($i.e.$, the coarse graining bulk contains 8 sites) as discussed in the previous section, because the plaquette VBS state (4 site per unit cell) is one of the candidate ground states at a zero field.
Apparently, the network adopted has suitable structure for the wave function of most candidates, namely the AF state, the OD state, and the VBS state.
Unfortunately the ER structure we use has difficulties dealing with incommensurate spin density wave (SDW) state, as common for most numerical method based on a finite unitcell with periodic boundary conditions. 
%\comment{Kenji: Is the following sentence necessary?}
If we can apply mutiple-level ERs on the SSL, the incommensurate SDW state may be captured approximately as in ref.~\onlinecite{Harada}. However, due to very high computational cost, it is difficult in our ER.
To summarize, we believe that our result is not much affected by the bias introduced from tensor network structure away from the critical region, although we can not rule out the possibility of the incommensurate SDW phase.

Secondly, the ER calculation we performed is essentially a variational approach. 
A well known feature of variational calculation is that optimization may converge into local minimums, instead of the global minimum - the true ground state.
In terms of energy landscape, the OD phase and the plaquette phase (which is fourfold degenerate on the square lattice) correspond to far separated, sharp and stable local minimums.
In other word, near the critical region, optimization calculations may end in both candidate states with finite possibilities, and it is impossible to ``tunnel`` between two different phases whence the optimization has reached the local minimum.
Hence, this is also a strong indication that the phase transition between the OD order and the plaquette VBS order is of first order.
A jump in order parameters and a sudden change of slope in energy can be observed in Fig.~\ref{ground-state}. 
In contrast, in the critical region between the VBS and AF phase, the two local minimums corresponding to the respective phases are shallow, not well separated and show tendency of merging with each other at the critical point.
Therefore the slope of energy does not change drastically, while $D$ and $M$ evolve smoothly across the phase boundary.
This is an indication of a second order phase transition, but we cannot rule out the possibility of cross over because of the severe finite-size dependence.

%%%%%%%%%%%%%%%%%%%%%%%%%%%%%%%%%%%
\section{Magnetic properties of SSL model}
\label{sec:MP}
%%%%%%%%%%%%%%%%%%%%%%%%%%%%%%%%%%

What makes the SSL model and its realization \SRCUBO fascinating is its magnetic property in a magnetic field.
It is well known from experiments that the coupling ratio is $J'/J\sim0.65$, where the ground state is given to be OD state (product of singlets).
In the presence of an external magnetic field, polarized triplets with $S_z=1$ are created, which can be treated as hardcore bosons moving on the effective square lattice.
Due to strong frustration in the lattice, single triple hopping is strongly suppressed, and effective repulsion between triplets governs the physics in the system when the coupling ratio satisfies $J'/J<0.5$.
Naturally, magnetic excitations form long range order and then, the system is in a magnetic plateau: an incompressible Mott phase.

Pioneer theoretical works confirmed several magnetic plateaux.
Among them, $1/3$ and $1/2$ are most well accepted, while plateaux in lower magnetization are still less clear. 
Recently, Sebastian {\it et.al.}\cite{Sebastian} have suggested $1/9$, $1/6$ plateaux and the other exotic ones at fractionalized magnetization from the quantum Hall analogy. 
Other theoretical works\cite{Ueda,Mila-1-8,Mila-plateaux,Capponi} also successfully confirmed the existence of these plateaux based on perturbative approaches from small $J'/J$ limit.
The result of our calculation is largely consistent with previous studies, but we have to highlight the following point: our calculation is not restricted to the small $J'/J$ limit, since we do {\it not} treat $J'$ interaction as a perturbation of the singlet product state.
As a matter of fact, we simply input our hamiltonian in the original square lattice, as shown in Fig.~\ref{mera-structure}, and the magnetic plateaux are obtained as the ground state of our pure variational calculation.
As a result, besides small $J'/J$ ratio ($< 0.5$), we also investigated large coupling ratios close to the phase transition point (into the plaquette VBS order).

%--------------------------------------------------------------------------
\begin{figure}
\centerline{\includegraphics[width=7.5cm,clip]{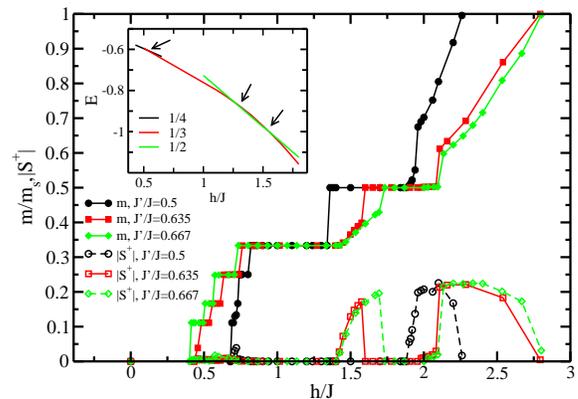}}
\caption{Magnetization of the system in response to external field $h$. Results obtained at three
different coupling ratio $J'/J=0.5, 0.635, 0.667$ are shown.
$1/2$, $1/3$ and $1/4$ plateaux are clearly shown. $1/9$, $1/8$ and $1/6$ plateaux are also observed.
Open symbols denote volume averaged off-diagonal spin component $\frac{1}{N}\sum \langle S^+ \rangle$ at
three coupling ratios.
%\comment{Kenji: could you add the value of $J'/J$ for red and green open symbols?}
They are indications to the supersolid phase.
Inset: The energies of variational local minimum branches that corresponds to 
states with $1/2$, $1/3$ and $1/4$ plateaus' dimer structures.
Cross points between two curves stand for jumps of magnetization.
%\comment{Jie: can you help me add some space between figures?}
}
\label{magnetic-plateaux}
\end{figure}
%--------------------------------------------------------------------------

%%%%%%%%%%%%%%%%%%%%%%%%%%%%%%%%%%%
\subsection{Magnetic plateaux}
%%%%%%%%%%%%%%%%%%%%%%%%%%%%%%%%%%

\begin{figure}
%\centerline{\includegraphics[width=7cm,clip]{figures/1_6.eps}}
\centerline{
\includegraphics[width=4cm,clip]{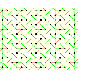}
\includegraphics[width=4cm,clip]{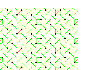}
}
\centerline{
\includegraphics[width=4cm,clip]{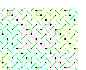}
\includegraphics[width=4cm,clip]{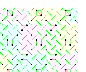}
}
\caption{Two sites spin-spin correlations $\langle S_i \cdot S_j \rangle$ for $J$ and $J'$ bonds plotted on the square lattice.
From red color to green, the correlation turns from ferromagnetic ({\it i.e.}, triplets) into AF (singlets).
The width of each plotted bond is proportional to the magnitude of correlation.
({\it i.e}, very thin line and dim color represents near-$0$ correlations.)
Open/closed circle at each site represents polarized/anti-parallel local spin, 
where radius of each circle stands for magnitude of $\langle S_z \rangle$.
Shadow region in the background shows the arrangement of unitcells.\\
Top two panels: correlation patterns obtained in $1/4$ and $1/8$ plateau states.
Bottom two panels: patterns obtained in $1/6$ and $1/9$ plateau states.
Large super-unitcells, which consist of one unitcell per category 
(represented by different color), 
are used to accommodate plateau states with small fraction numbers.
}
\label{pattern}
\end{figure}

\begin{figure}
\centerline{
\includegraphics[width=4cm,clip]{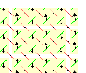}
\includegraphics[width=4cm,clip]{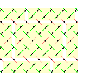}
}
\centerline{
\includegraphics[width=4cm,clip]{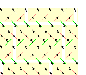}
\includegraphics[width=4cm,clip]{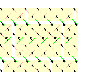}
}
\centerline{}
\caption{Dimer patterns obtained for $1/2$, $1/3$ (top two panels) and P3 (bottom two panels) supersolid states.
The width of each dimer is proportional to the magnitude of correlations between two spins.
From AF to ferromagnetic correlations, the color of each dimer can change from green to red continuously.
Arrow on the each dimer stands for the in-plane component of the coherent vector $(\Phi, \Theta)$,
and the radius of circle on each dimer is proportional to the vector component normal to the plane.
}
%Panel (h): Illustration of plaquette VBS order for $J'/J>0.685$.}
\label{pattern-ss}
\end{figure}

It is clear judging from Fig.~\ref{magnetic-plateaux}, that we obtain well accepted $1/4$, $1/3$ and $1/2$ plateaux in the magnetic curve when external field is applied.
In addition to these three plateaus, minor plateaux such as $1/6, 1/8$ and $1/9$ are observed as well
In Fig.~\ref{pattern}, 
we plotted correlations between neighboring spin pairs $\langle \cdot \rangle$ (for both $J$ and $J'$ bonds), as well as $\langle S_z \rangle $ for all local spins, for minor plateaux 
(similar information for $1/2$ and $1/3$ can be read in Fig.~\ref{pattern-ss}).
These dimer patterns are consistent with previous studies.\cite{Momoi,Mila-plateaux,Takigawa,Ueda}
Notice that color and width of each dimer we draw stand for spin correlations between two sites forming the dimer, where red and green stand for ferromagnetic (triplet) and AF (singlet) correlation, respectively. The magnitude of $S_z$ for each local spin is also plotted as the radius of circle on each site.

It is worth mentioning that the 1/8 plateau appears only around $J'/J=0.5$ and the local moment distribution $\langle S^{z}_{i}\rangle$ is of the so called rhomboid type, although the energy difference between rhomboid type and square type is extremely small. 
%good agreement with one reported in previous NMR measurements,\cite{Takigawa} namely rhomboid type. 
%\textcolor{blue}{After discussion with Prof. Takigawa...} {\bf Note that the spin configration in the 1/8 plateau has been corrected to be square type.\cite{Recent_NMR}}
%\comment{Jie: The energies of Rhomboid type and square type are very close, down to 4 or 5th digit.}
%In \SRCUBO, the coupling ratio was estimated as $J'/J \sim 0.65$.\cite{Ueda} 
For $J'/J\agt 0.5$, we find that the 1/8 plateau state (rhomboid type configuration) has higher energy compared with other plateaux. 
This discrepancy may arise from further long-range interactions. 

Another interesting point is that the 2/9 plateau state, proposed in ref.~\onlinecite{Mila-plateaux},
which can be covered by our super-unitcell tensor network wave function (and emerge as local minima in our variational calculation), turns out to be worse in energy compared to other low magnetization plateau states (1/4,1/6) in relevant parameter region.
This fact indicates that the PCUT treatment \cite{Mila-plateaux} seems to be already broken down for $J'/J>0.5$.

Again here we discuss the validity of our variational ground states.
It is obvious that plateau states with different fraction number correspond to local minimums, in terms of energy landscape in our calculation.
In the presence of relatively large external field. ($i.e.$, $m/m_{s}\ge 1/3$, 
while $m_{s}$ in the saturation magnetization), we find there are two major groups of local minimums: one corresponds $1/2$ plateau and the other is $1/3$.
These two local minimums are stable, sharp and far separated, which means tunneling from one plateau state to another is nearly impossible to happen, especially at the later part of the optimization.
The reason is simply that these two plateau states are characterized by different periodicity.
We use the existence of stable local minimums to our advantage.
We can obtain energy of a certain plateau state as a monotonic function of field $h$, if we choose to load tensors from previous calculations (that is know to host the plateau state) instead of starting from random initial condition.
Phase boundaries are then determined by the crossing point of energy curves of different plateau states, as shown in inset Fig.~\ref{magnetic-plateaux}.
Each crossing point corresponds to a jump in the magnetic curve, which is a strong first-order phase transition. Similar techniques were also used in other variational method as well.\cite{Mila-plateaux}
However, in lower magnetizations ($m\le 1/4$), the energy gap between different plateau states are fairly close and the phase boundaries between different plateau states are less clear, because the energy for each plateau state is sensitive to changing of unitcell geometry and other factors.
It is also necessary to point out that the dimension $\chi$ of the coarse grained system in our calculation has very limited impact on results of magnetic plateaux. 
In fact, all measured observables are not much dependent on $\chi$, and the magnetization 
tends to reach the correct fraction number even at smallest dimension $\chi=2$.

%--------------------------------------------------------------------------
%--------------------------------------------------------------------------

%--------------------------------------------------------------------------
%--------------------------------------------------------------------------
%
\subsection{Supersolid states}

In addition to multi plateaux, it was predicted that spin supersolid phases appear in the SSL model when magnetic fields exists.\cite{Momoi,Mila-supersolid}
It has been pointed out that correlated hopping of triplets is crucial to stabilize the supersolid phases.
The correlated hopping ({\it i.e.}, triplets, which are regarded as bosons, hop from one dimer to another when a neighboring boson is present) contributes as the main source of kinetic energy in the system, in contrast to single triplet hopping which is largely suppressed by frustration.
Such correlated hopping can lead to supersolid phase, where diagonal (crystalline solid) and off-diagonal (superfluidity) long range order coexist.\cite{supersolid}
For the SSL model with considerably large coupling ratio $J'/J$, supersolid phases are predicted to appear above $1/3$ and $1/2$ plateaus in the high field region.
 
Quite recently, high field measurements on \SRCUBO over 100 [T] have been achieved and the presence of the 1/2 plateau has been confirmed.\cite{Jaime} 
In real compound, the Dzyaloshinski-Moriya (DM) interaction, which breaks U(1) symmetry extrinsically, can not be ignored, therefore superfluid transition is not prohibited in the compound, when the temperature decreases. 
Fortunately, the magnitude of the DM interaction in \SRCUBO was estimated as $D_{\rm DM}\sim 2{\rm [K]}$\cite{Ueda} and it is about 0.2\% of the strongest coupling $J$. 
This allows us to expect that the weak DM interaction would fix the phase of off-diagonal components.
Note that the transition related with superfuildity is not recovered due to the intrinsically symmetry of spin Hamiltonian.
Beside this point, theoretically, the accuracy of previous studies' evaluation of correlated hopping effect in large $J'/J$ region lacks,\cite{Capponi}
 and to the best of our knowledge, supersolid phase are not clearly observed in unbiased numerical studies for the SSL model. 

From Fig.~\ref{magnetic-plateaux}, it is clear that 
the magnetization $m$ respond to the external 
field continuously (gapless mode) just above the $1/3$ and $1/2$ plateaux, 
%for relatively large coupling ratio $J'/J=0.635$ and $0.667$,
strongly indicating existence of supersolid phases.
%On the other hand, no such behavior is observed for $J'/J=0.5$.
To further support this idea, we notice
that all dimers in our system can be considered as a mixing of singlet state and triplet excitation.
Using a classical coherent vector ($\Omega =cos\Phi sin\Theta , sin\Phi sin\Theta , cos\Phi$), 
we can represent a dimer state as:
\eqnarray 
|\Phi , \Theta \rangle = e^{i\frac{\Phi}{2}} \sin \left(\frac{\Theta}{2}\right) | s \rangle
+ e^{-i\frac{\Phi}{2}} \cos \left(\frac{\Theta}{2} \right) | t \rangle,
\label{coherent}
\endeqnarray
where $|t\rangle$ and $|s\rangle$ stand for triplet and singlet states, respectively.
Only when superfluidity of triplet excitations exist in the system, $\Theta$ can take angles
not equal to $0,\pi$.
It is easy to show that under such circumstance, the off-diagonal components of original spins
in the lattice take non zero values as:
\eqnarray 
\langle S_1^{+} \rangle = + \frac{1}{2\sqrt{2}} \sin \Theta,\quad
\langle S_2^{+} \rangle = - \frac{1}{2\sqrt{2}} \sin \Theta.
\endeqnarray
As a result, off-diagonal elements are anti-parallel for the two sites $1$ and $2$ on a dimer bond.

We measure off-diagonal elements of original spin $|S^+|$ as the order parameter of supersolid phase.
Results are shown in Fig.~\ref{magnetic-plateaux}, indicating presence of strong supersolid phase above $1/3$ (for large coupling ratio $J'/J=0.635$ and $0.667$) and $1/2$ (for all $J'/J$ parameters we tested) plateaux.  
Notice that the $1/3$ supersolid phase appears only in large coupling ratio region, due to the fact that
for large $J'/J$, the repulsion between dimers are largely reduced, while related correlated hopping remains finite.
On the other hand, for relative small $J'/J=0.5$, the $1/3$ supersolid does not appear as the ground state
in our calculation.

The geometry of crystalline pattern of the triplets plays an important role in the correlated hopping process,
as pointed out by several studies.\cite{Ueda,Mila-supersolid,Mila-plateaux}
More specifically, for both $1/3$ and $1/2$ plateau states, triplet excitations (and the remaining singlet dimers)
form stripes in $(\pi,\pi)$
direction of the SSL, which corresponds to horizontal and vertical lines in our illustrations 
(Fig.~\ref{pattern-ss}). 
In the supersolid phase, additional triplets can hop both parallel to the 
stripes ({\it i.e.}, flow inside the ``canal'' between two triplet stripes)
and perpendicular to it (``jump'' across the stripe).
It has been pointed out that hopping matrices with regard to hopping 
inside and across a stripe of triplet are of opposite sign.
As a result, the superfluid components (presented by angle $\Phi$ in (\ref{coherent}))
is parallel inside stripes, and anti-parallel across stripes.

Using the local density matrix obtained from MERA calculation, 
we are able to calculate the superfluid component for each local dimer, as
\eqnarray 
Z&=&|t\rangle \langle t| - |s \rangle \langle s|,\\
X&=&|t\rangle \langle s| + |s \rangle \langle t|,\\
Y&=&i(|t\rangle \langle s| - |s\rangle \langle t|).
\endeqnarray
Resulted coherent vectors are plotted in Fig.~\ref{pattern-ss},
where arrows show the direction of superfluid in the XY plane.
%and radius stands for the magnitude of diagonal Z component.
Our result is consistent with the expectation for both $1/2$ and $1/3$ supersolid phases.
Notice that although supersolid vectors form LRO, the direction of such ordered ``magnetization'' itself is not fixed, which can turn out to be any angle within the plane, since random tensors are used as the starting point of our variational calculation.
Due to the ordering of superfluid component, the spatial symmetry in the supersolid phase is different compared with respective plateau order.
One need to be cautious setting up the unitcell for numerical simulations of supersolid phases.
   
%%%%%%%%%%%%%%%%%%%%%%%%%%%%%%%%%%%
%\section{phase diagram}
%%%%%%%%%%%%%%%%%%%%%%%%%%%%%%%%%%
%--------------------------------------------------------------------------
\begin{figure}
\centerline{\includegraphics[width=7cm,clip]{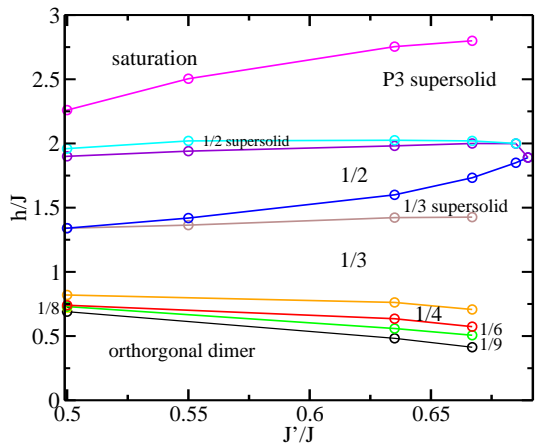}}
\caption{We propose a tentative phase diagram focused on large coupling ratio region $J'/J \geq 0.5$.
The $1/2$ plateaux persist to the largest coupling ratio $J'/J=0.667$ 
(close to the transition point to the plaquette VBS phase) we have examined.
$1/8$ plateau only appear at $J'/J=0.5$.
Supersolid phase exist just above the $1/3$ and $1/2$ plateaux at large coupling ratio.
Above $1/2$ supersolid, another supersolid phase with period $3$ emerge as ground state.
}
\label{phase-diagram}
\end{figure}
%--------------------------------------------------------------------------

Based on these results, we provide a schematic phase diagram in Fig.~\ref{phase-diagram}, with coupling ratio $J'/J$ and external field $h$ as parameters.
At $J'/J=0.5$, the phase separation point from our calculation is in good agreement with the phase diagram from Capponi {\it et.al.}.\cite{Capponi}
Notice that $2/9$ plateau is not well established in our calculation, as the energy of such LRO pattern is not favored in energy (although they do appear as local minimum
in optimization).

Since our method is not limited to small coupling ratio limit, we mainly probed large $J'/J$ region.
Our result largely agree with the schematic phase diagram proposed in strong-coupling
expansion calculation by Momoi {\it et.al.}\cite{Momoi} with a few discrepancies.

%%%%%% until here 2012.10.24. T. Suzuki

%need to change this part%
First of all, the external field region $h$ where the $1/2$ plateau exists does decrease significantly
when the coupling ratio $J'/J$ is increased, but {\it persist} until the largest ratio
we tested, $J'/J=0.687$,  which is the phase separation point to the OD order
in the original SSL model with no external field.
The main reason for the shrinking of $1/2$ plateau has been revealed in previous studies:
the strength of the nearest neighbor repulsion between dimers significantly reduces 
in large $J'/J$ region.
It is however not clear whether the $1/2$ plateau shall remain in the compound \SRCUBO
(with $J'/J \approx 0.635$), or materials with even larger coupling ratios.
According to our ER calculation, the $1/2$ plateau still gains its foot in the competition against
supersolid phases at large coupling region.

Secondly, the $1/3$ supersolid phase does not appear at $J'/J=0.5$, rather, 
emerge at fairly large coupling ratio. This is slightly inconsistent with Momoi's prediction, 
but agrees with more carefully executed perturbation calculations.\cite{Capponi}
%This is consistent with more detailed calculations.\cite{Capponi}

%{\bf Jie: The following part I am not certain to include:
%We want to point out that when the coupling ratio $J'/J$ is increased, 
%$1/6$ and $1/9$ plateaus start to gain advantages in energy compared with $1/8$
%plateau. As a result, the $1/8$ plateau only emerge as the ground state at relatively
%small coupling ratio $J'/J=0.5$.
%}

Above the $1/2$ supersolid phase, through a frist order phase transition,
%\comment{Jie:edited according to Kenji's question} 
the system enters another supersolid phase (with different
triplet pattern), in which its magnetization keeps responding continuously to external field,
and $\langle S^{+} \rangle$ remains finite, as shown in Fig.~\ref{magnetic-plateaux}.
In contrast to early studies, this supersolid phase emerges in our calculation and features a crystalline pattern of singlet-triplet description.
As a matter of fact, it can be considered as a natural extension to the $1/3$ plateau state, as the period remains to be 3 rows of dimers.
More specifically, when field $h$ is large enough, while one stripe of triplet remain unchanged, two remaining stripes of singlets start to covert to triplets, one row follows another, until all dimers are polarized.
The conversion is in a continuous fashion, {\it i.e.}, the correlations $\langle S_iS_j \rangle$ on the dimer change from ferromagnetic to AF smoothly. 
(See example in Fig.~\ref{pattern-ss}.)
We refer to such supersolid phase as P3 supersolid.
%\comment{Kenji: the period of P3 supersolid is different from that of 1/2 one. Is there a transition between them when a magnetic field increases? The change of sloop in a magnetizaion curve is an indication of such transiton?
%Jie: yes, there is a first order phase transition. You can observe a jump from 1/2 SS to 1/3 SS}
The reason that the P3 supersolid phase is favored in energy compared to $1/2$ supersolid at large magnetization region is unclear. 
One possible reason is that during the smooth polarization from low magnetization to full saturation,
({\it i.e.}, $1/3-2/3-1$), the diagonal LRO is largely unaffected, which features a period of 3 rows
of dimers.
%\comment{Takafumi: I corrected here. Please check meanings you wanted to mension.} 
Correlated hopping may also be easier in P3 crystalline patterns, which render P3 supersolid an advantage in kinetic energy.
%\comment{Jie: Need some careful analysis of correlated hopping. Maybe we can strictly prove it?}

\section{Summary}
Using MERA method, we are able to calculate the ground state of the SSL model both in the absence and presence of external magnetic field.
Our calculation does not make any approximation of the Hamiltonian itself, unlike some early studies, and the ground state we obtained is the result of pure variational calculation.
Thermodynamic limit has been reached by meanfield-like approximations (FRMERA, although there is subtle difference).
We have showed clear evidence of a plaquette VBS order intermediating the AF and the decoupled dimer phase.
We have found various magnetic plateaux in the presence of external field, including $1/2, 1/3, 1/4, 1/6, 1/8, 1/9$.
We have also confirmed the existence of supersolid phases above $1/3$ and $1/2$ plateaux, 
where superfluidity (off-diagonal LRO) and crystalline pattern of triplet excitations coexist.
Above the $1/2$ supersolid, another supersolid phase which features period 3 dimer lattice structure persists all the way to saturation.

We have to point out that the FRMERA method faces difficulties in dealing with incommensurate phases, due to finite unitcell,
and lacks in description of long range entanglement that is far beyond the size of the unitcell.
Nevertheless, it is a very effective numerical method for studying gapped quantum spin systems with frustration.
%\comment{Jie: need to decide if this part is necessary}
We have to point out here that DM interaction, which is believed
to have profound impact in the material's magnetic properties, is regrettably not included in our discussion. 
Also effect of higher-order-neighbor interactions have been not included either\cite{Suzuki}.
We notice that it is practical to include them in the calculation,
if our MERA tensor network is modified accordingly in order to restrain the cost.
We choose to leave these possibilities to future studies.  
%}
%\comment{Kenji:I think that this part necessary.}

%Acknowledgement:
%We would like to thank ...
This research was supported in part by Grants-in-Aid for Scientific Research No. 23540450.
Calculations are performed on the Kashiwa super-computer located in ISSP, university of Tokyo.
%\comment{Kenji: My grant aid.}

\end{document}